\documentclass[10pt,prl,aps,twocolumn,groupedaddress]{revtex4}
\usepackage{amssymb,amsmath,amsfonts,graphicx,latexsym,color}

\usepackage[latin9]{inputenc} 
\usepackage{color}
\usepackage{amsmath}
\usepackage{amssymb}
\usepackage{graphicx} 
\usepackage{ulem}

\usepackage{times}
\usepackage{mathptmx}

\usepackage[]{hyperref}

\begin{document}
 
\newcommand{\bra}[1]{\left< #1 \right|}
\newcommand{\ket}[1]{\left| #1 \right>}
\newcommand{\heff}{\hat{H}_{\mathrm{eff}}}
\newcommand{\tred}[1]{{\color{red} {#1}}}
\newcommand{\tblue}[1]{{\color{blue} #1}}
\newcommand{\eps}{\epsilon}
\newcommand{\mI}[1]{\hat{\mathcal{I}}_{#1}}
\newcommand{\U}[2]{\hat{U}(#1,#2)}
\newcommand{\Ud}[2]{\hat{U}^\dagger(#1,#2)}
\newcommand{\mA}{\hat{\mathcal{A}}}
\newcommand{\tc}{\tilde{a}}
\newcommand{\tb}{\tilde{b}}
\newcommand{\hb}{\hat{b}}
\newcommand{\hbd}{\hat{b}^\dagger}
\newcommand{\hc}{\tc}
\newcommand{\ha}{\hat{a}}
\newcommand{\hcd}{\hc^\dagger}
\newcommand{\hh}{\hat H}
\newcommand{\mb}{}
\newcommand{\cN}{\mathcal{N}}
\newcommand{\cM}{\mathcal{M}}
\newcommand{\hs}{\hat{\sigma}}

\title{Periodic thermodynamics of isolated quantum systems}

\author{Achilleas Lazarides$^{1}$, Arnab Das$^{1,2}$ and Roderich Moessner$^{1}$}
\affiliation{$^{1}$ Max-Planck-Institut f\"ur Physik komplexer Systeme, 01187 Dresden, Germany}
\affiliation{$^{2}$ Theoretical Physics Department, Indian Association for the Cultivation of Science, Kolkata 700032, India}

\date{\today}

\begin{abstract}
The nature of the behaviour of an isolated many-body quantum system 
periodically driven in time
has been an open question since the beginning of quantum 
mechanics~\cite{shirley1965solution,Sambe,Hanggi-Rev,ArnabFreezing,Eckardt2005a,floquetTopo}. 
After an initial transient, such a system is known to synchronize with the driving; 
in contrast to the non-driven case, no fundamental principle has been proposed for 
constructing the resulting non-equilibrium state. 
Here, we analytically show that, for a class of integrable systems, 
the relevant ensemble is constructed by maximizing an appropriately 
defined entropy subject to 
constraints~\cite{Jaynes1957a} which we explicitly
identify. This result constitutes a generalisation of the
concepts of equilibrium statistical mechanics to a class of
far-from-equilibrium-systems, up to now mainly 
accessible using ad-hoc methods.
\end{abstract}

\maketitle

There has recently been significant progress in our understanding of 
statistical mechanics based on the twin concepts of 
\emph{equilibration}, the approach of a large, closed system's state to some steady 
state~\cite{Srednicki1994,Deutsch1991,Rigol2008b,Cazalilla2012a,Rigol2007a,
Calabrese2011a,Reimann2008a,Fagotti2013},
as well as of \emph{thermalization}, 
when this steady state depends only upon a small number of quantities. 
Starting from ideas due to 
Jaynes~\cite{Jaynes1957a}, Srednicki and Deutsch~\cite{Srednicki1994,Deutsch1991} 
and Popescu {\it et al}~\cite{Popescu2006}, 
both integrable and non-integrable closed, non-driven many-body systems have 
thus been shown to thermalize~\cite{Fagotti2013,Rigol2007a,Rigol2008b}.

On the other hand, the study of periodically driven systems has also 
had a long history. Following early foundational work by
Shirley~\cite{shirley1965solution} and Sambe~\cite{Sambe}, 
substantial theoretical and experimental
progress has been made recently~\cite{Eckardt2005a,Hanggi-Rev,ArnabFreezing,floquetTopo,StoferleModulation,ArimondoBreathingTrap,Haller2009a,ChenModulation,Iadecola}. 

Here, we combine ideas from the two areas to extend the concept
of thermalization to the out-of-equilibrium  
case of periodically driven systems. 
By devising a mapping of the system
to a set of effectively non-driven systems we show 
that a periodically driven system asymptotically approaches a 
time-periodic steady state at long times 
(see, e.g., \cite{Russomanno2012} and our Suppl. Mat.). 
Specializing to a large class of integrable systems,
we analytically show that Jaynes' entropy maximisation 
principle~\cite{Jaynes1957a} gives a statistical 
mechanical description of the long-time, synchronized dynamics for
infinite systems, and study the approach to this equilibrium state as
a function of both the system size and time. Finally,
we explain how our proposed setup is achievable with current
experimental techniques.

\begin{figure*}[t]
  \begin{center}
    \includegraphics[width=17cm]{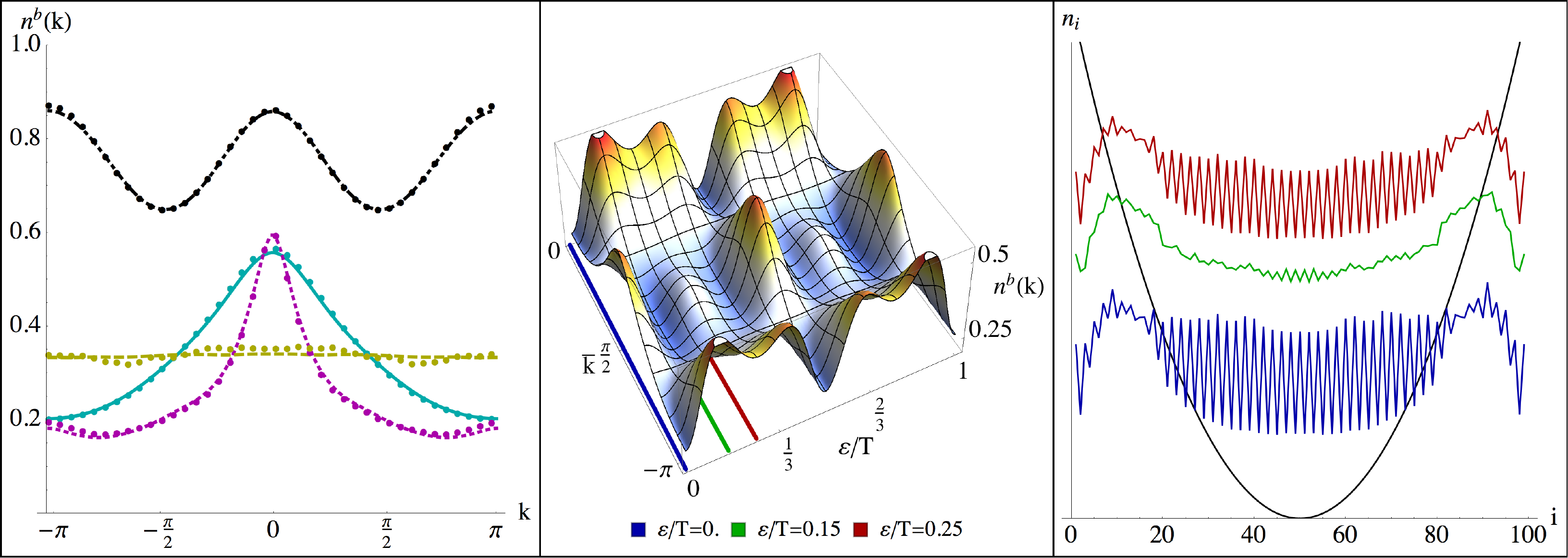}
  \end{center}
  \caption{
  Characterisation of the synchronised steady state.
        {\bf Left}: Stroboscopic momentum distribution, 
        $\hat{n}(k)=L^{-1} \sum_{i,j}\hb^\dagger_i\hb_j
          \exp(-2\pi i k(i-j)L^{-1})$, demonstrating the wide range of behaviour that 
        occurs for varying parameters. The points correspond to snapshots
		of the dynamical evolution at late times ($t=490T$) for $L=200$,
		while the continuous lines correspond to the PGE prediction.
		From top to bottom at the extreme left end
        of the plot, the amplitude of the superlattice potential, 
        frequency and filling factor,
        $(\Delta,\delta J,\omega,\nu)$ are $(0.6,0.5,1.6,3/4)$ (black,
        dot-dashed), $(4,0.5,1.5,1/3)$ (yellow, dashed), $(4,0.75,2,1/3)$ (cyan,
        full), $(0.6,0.5,2,1/4)$ (magenta, dotted) and $\eps=0$. 
		The next two panels correspond to 
        the parameters for the cyan full line. 
		{\bf Centre}: 
		Expectation value of the momentum distribution $\hat n(k)$ of the bosons during a
        single period in the synchronized state as a function of the time in the 
        period, $\epsilon$. The three lines on the time-momentum plane indicate the 
        times $\epsilon/T=0, 
        0.15, 0.25$ for which density distributions are shown in the rightmost panel. 
        The momentum distribution undergoes qualitative changes: at some points of the
        period it has a single maximum at $k=0$ while at others it acquires double maxima
        at the edges of the Brillouin zone.
        {\bf Right}: Each trace shows the expectation value of the 
        density of the bosons, $\hat{n}^b_i=\hb^\dagger_i\hb_i$, at the time indicated in
		the middle panel by the line of the 
        same colour, for a lattice size $L=100$ and offset for better visibility. 
        The black line indicates the time average of the applied potential; the
        density peaks at the edges despite the potential being highest there, 
        indicating a strongly non-equilibrium situation.
  \label{fig:final-Values-And-Period}
}
\end{figure*}

\emph{Synchronization--} The starting point for our analysis is the
synchronization of the system with the driving, which may be seen as follows.

Consider a time-periodic Hamiltonian 
 $\hh\mb(t)=\hh\mb(t+T)$ and 
denote the time evolution operator over a period starting from time $0\leq \epsilon<T$ 
by $\hat U\mb\left(\epsilon,\epsilon+T\right).$ 
Taking $\hbar = 1$, we define an effective Hamiltonian $\heff$ via
\begin{equation}
   \exp{\left[-i {\heff}\mb T\right]}=\hat U\mb\left(0,T\right),
  \label{eq:effective-H}
\end{equation}
$\heff$ is a time-independent effective Hamiltonian which takes an initial state
at $t=0$ to the same final state at $t=T$ as the real time-dependent
Hamiltonian $\hh\mb(t)$. 

We concentrate on ``stroboscopic'' observations, that is, observations at discrete
points of time separated by a period, $t_n = \epsilon + nT$ for a given
$\epsilon$. The expectation value of an arbitrary time-independent operator $\mathcal{\hat O}$ at time $t$, $\mathcal{O}(t)=\left<\psi(t)\right|\mathcal{\hat O}\left|\psi(t)\right>$, is
\begin{equation}
  \mathcal{O}(t_n)=\bra{\psi(0)} 
    \mathrm{e}^{i\heff\mb nT}
    \mathcal{\hat O^{(\epsilon)}}
    \mathrm{e}^{-i\heff\mb nT}
    \ket{\psi(0)} 
  \label{eq:expectation-O}
\end{equation}
where $\hat{\mathcal{O}}^{(\epsilon)}
  =\hat{U}^\dagger\mb(0,\epsilon)\hat{\mathcal{O}}\hat{U}\mb(0,\epsilon).$
We have thus recast the time evolution into evolution under 
a time-independent Hamiltonian, at the price of introducing a set of new
operators $\hat{\mathcal{O}}^{(\epsilon)}$.

By analogy to a static quench~\cite{Reimann2008a,Cazalilla2012a} 
(see Supplementary Material for a discussion of the necessary conditions), 
one can show that each series $\{{\mathcal
O}(t_n); \quad n = 0,1,2\ldots\}$ converges to a fixed value. This immediately implies
that the long-time behaviour of the system is periodic in time, i.~e., synchronised.

\emph{Construction of the periodic ensemble--}
We now come to the main part of our work where we show that 
Jaynes' idea of entropy 
maximization~\cite{Jaynes1957a,Rigol2007a,Cazalilla2012a} 
remains valid away from equilibrium for this class of models.
In order to demonstrate that this is correct, 
we restrict ourselves to a class of tractable integrable Hamiltonians. 
For infinite systems, we show \emph{analytically} that this 
ensemble correctly reproduces all correlation functions. 
For finite systems, we study the approach to the thermodynamic limit 
in a spatially inhomogeneous system of hard-core 
bosons (HCBs). 

The Hamiltonians we consider are of the form
\begin{equation}
  \hat{H}(t) = \sum_i\left[
      \ha_i^\dagger \cM_{i,j}(t)\ha_j + 
      \ha_i^\dagger \cN_{i,j}(t) \ha_j^\dagger + \mathrm{h.c.}
    \right],
  \label{eq:quadratic}
\end{equation}
with the $\ha_i$ fermionic or bosonic operators, 
$\left[a_i,a^\dagger_j\right]_{\pm}=\delta_{i,j}$, and $\cM,\cN$ are
complex matrices. 
In cases of interest, the nonlinear, nonlocal
transformation that brings the physical Hamiltonian to this form maps
local observables to highly nonlocal, nonlinear functions of the
$\ha$ operators. 

For Hamiltonians bilinear in the operators $\ha$, 
$\heff$ are bilinear and may therefore 
be brought to the form
\begin{equation}
  \heff=\sum_{p=1}^L \omega_p \tc^\dagger_p\tc_p
  \label{eq:heff}
\end{equation}
by a unitary transformation ($L$ is the system size). The
operators $\hat{\mathcal{I}}_p(t):=\hat{U}(0,t)\tc^\dagger_p\tc_p\hat{U}^\dagger(0,t)$ 
(of which there are $L$) correspond to conserved quantities,
$\bra{\psi(t)}\hat{\mathcal{I}}_p(t)\ket{\psi(t)}
  =\bra{\psi(0)}\hat{\mathcal{I}}_p(0)\ket{\psi(0)}$ for all $t$, and are temporally 
periodic.

%

We now describe how to obtain the 
statistical ensemble describing the long-time behaviour of this
system after a number of periods have elapsed. 
Given the set
$\left\{\hat{\mathcal{I}}_p(t)\right\}$ 
we construct the most general distribution maximizing
Shannon's entropy in the space of periodic operators, subject to the constraints
given by the conservation laws.
The resulting ``periodic Gibbs ensemble'' (PGE) density operator is
\begin{equation}
  \label{eq:pge-rho}
  \hat\rho_{PGE}(t)=\mathcal{Z}^{-1}\exp\left(-\sum_{p}\lambda_{p}\hat{\mathcal{I}}_{p}(t)\right)
\end{equation}
\noindent
with the $\lambda_p$ fixed by requiring that 
$\bra{\psi(0)}\hat{\mathcal{I}}_p(0)\ket{\psi(0)}=\mathrm{tr}\left( \hat\rho_{PGE}(0) 
\hat{\mathcal{I}}_p(0)\right)$ and $\mathcal{Z}=\left(\mathrm{tr}\,\hat\rho_{PGE}(t)\right)^{-1}$ 
a (time-independent) normalization factor. 

Operator $\hat{\rho}_{PGE}(t)$ has the following two properties: First, it correctly gives the conserved quantities:
$\mathrm{tr}\left(\hc^\dagger_p\hc_q\hat{\rho}_{PGE}(t)\right)=\delta_{p,q}\bra{\psi(t)}
  \hat{\mathcal{I}}_p(t)\ket{\psi(t)}$.
Secondly, since the $\mI{p}$ are periodic in time, it is itself manifestly periodic with time:  
$\hat\rho_{PGE}(t)=\hat\rho_{PGE}(t+T)$.

Finally we can analytically show that the PGE density matrix exactly reproduces all
correlation functions in the thermodynamic limit; this somewhat lengthy but ultimately
elementary calculation is described in the Supplementary Material. This constitutes
our central conceptual result.

\emph{Application to Finite Systems: Numerical Results --} Let us now supplement the above
exact and general results using numerical simulations for specific, finite systems.
While the proof for the correctness of the PGE 
is strictly applicable only in the thermodynamic limit, we shall
see that the deviation of finite systems from the PGE result rapidly decreases with system
size.

A number of different physical systems may be 
mapped to Eq.~\eqref{eq:quadratic} (see Supplementary Material). Here we present
numerical results for the experimentally relevant case of HCBs subject
to a simple potential, the Hamiltonian for which reads
\begin{equation}
  \hat{H}_{b}(t)=-\frac{1}{2}\sum_i J_i(t) \hbd_i \hb_{i+1}+\mathrm{h.c.}
    +\sum_i V_i(t)\hbd_i \hb_i
    \label{eq:hcb}
\end{equation}
with the $\hb_i$ HCBs.
The HCBs are described by operators $\hb$ obeying bosonic commutation relations, $\left[\hb_i,\hb_j^\dagger\right]=\delta_{i,j}$, with the addional hard-core
condition $\hb^2_i=0$.
A Jordan-Wigner transformation, $\hb_i=\ha_i \prod_{j<i}(-1)^{\hat{n}_j}$ with 
$\hat{n}_j=\hb_j^\dagger \hb_j=\ha_j^\dagger\ha_j$, maps
$\hat{H}_{b}(t)$ to Eq.~\eqref{eq:quadratic} with
$\cM_{i,j}(t)=-\frac{1}{2}J_i(t)\left(\delta_{i+1,j}+\delta_{i-1,j}\right)+\delta_{i,j}V_i(t)
$, $\cN_{i,j}=0$ and fermionic commutation relations for the $\ha$.

Here we focus on a time-dependent superlattice potential superposed on a
quadratic potential,
$V_i(t)=\frac{1}{2}\left(\left(i-L/2\right)/\ell_{ho}\right)^2 +\Delta
(-1)^i\cos\left(\omega t\right)$ and a time-dependent hopping amplitude $J_i(t)=J+\delta
J\cos(\omega t)$ with $\omega=2\pi/T$.
The protocol we use is to prepare the system in the ground state 
in the presence of a harmonic potential
$V^{(0)}_i=\frac{1}{2}\left(\left(i-L/2\right)/\ell_{ho}\right)^2$, fixing
$\ell_{ho}=N$. This allows us to take the thermodynamic limit, 
since for large total number of particles the dimensionless 
parameter~\cite{Rigol2004universal} $\tilde{\rho}=N_b/\ell_{ho}$ plays a role analogous to the density in 
the uniform limit. Results with different system sizes but constant $\tilde{\rho}$ are therefore
comparable.

At time $t=0$, the driving is switched on so that the total Hamiltonian is
$\hat{H}_{b}(t)=
  -\frac{1}{2}J\sum_i \hbd_i \hb_{i+1}+\mathrm{hc}+\sum_i V_i(t)\hbd_i \hb_i$
with $V_i(t)=V^{(0)}_i+\Delta (-1)^i\cos\left(2\pi t/T\right)$.

Concentrating on the experimentally accessible momentum distribution
of the bosons, $\hat{n}^{(b)}(k)=L^{-1} \sum_{i,j}\hb^\dagger_i\hb_j\exp(-2\pi
k(i-j)L^{-1})$ we use the numerical method used in, \emph{inter
alia},~\cite{rigol2005ModPhysLett}; it consists of solving the
fermionic time-dependent problem and, at the end, inverting the Jordan-Wigner 
transformation.\footnote{
It is worth pointing out that $\hat{n}^{(b)}$ for the bosons is neither 
bilinear nor local in terms of the Jordan-Wigner fermions, since $\hbd_i \hb_j=\ha_i^\dagger \left(\prod_{i\leq m<j}(-1)^{\hat{n}_m}\right) \ha_j$. We therefore expect the PGE 
predictions to
only approximate the real dynamics, becoming exact at the thermodynamic
limit.
}

We begin by demonstrating a number of possible periodic states, 
corresponding to different
parameters of the model. The leftmost panel of Figure~\ref{fig:final-Values-And-Period}
shows snapshots of the PGE momentum distribution 
$\mathrm{tr}\left(\hat{\rho}_{PGE}\hat{n}^{(b)}(k)\right)$
 at the beginning of each period
($\eps=0$) for different parameter values. We emphasise that,
away from the high-frequency regime, the corresponding
time-averaged Hamiltonian~\cite{Eckardt2005a,locDriven} is not an appropriate description.
As a striking example, the black line shows a momentum distribution with peaks at the edges
of the Brillouin zone. Concentrating now on the parameters corresponding to the cyan line,
the central panel shows the time evolution of the momentum distribution over an entire
period. Note, the system evolves through states in which the momentum is peaked 
at different locations of the Brillouin zone. 
Finally, the rightmost panel shows three snapshots of the
density distribution of the same system at times indicated by the coloured lines in the
central panel. The high spatial frequency oscillations and the peaking of the density at the
edges is also very different from what would be obtained had the system been well-described by a
time-averaged Hamiltonian, since the time-averaged potential (shown in black) is smooth and
its potential highest at the edges.

We next discuss the approach to the long-time periodic state as a function of time and system
sizes. After showing that the stroboscopic values of observables
approach, then oscillate around, a constant value for each $\eps$, we proceed to demonstrate
that both this average value and the relative magnitude of the oscillations away from it decay
to zero with increasing system size, in agreement with our analytical results for infinite
systems. The approach is rapid: within a few periods, the system is practically thermalized.

\begin{figure}[t]
  \begin{center}
    \includegraphics[width=8.5cm]{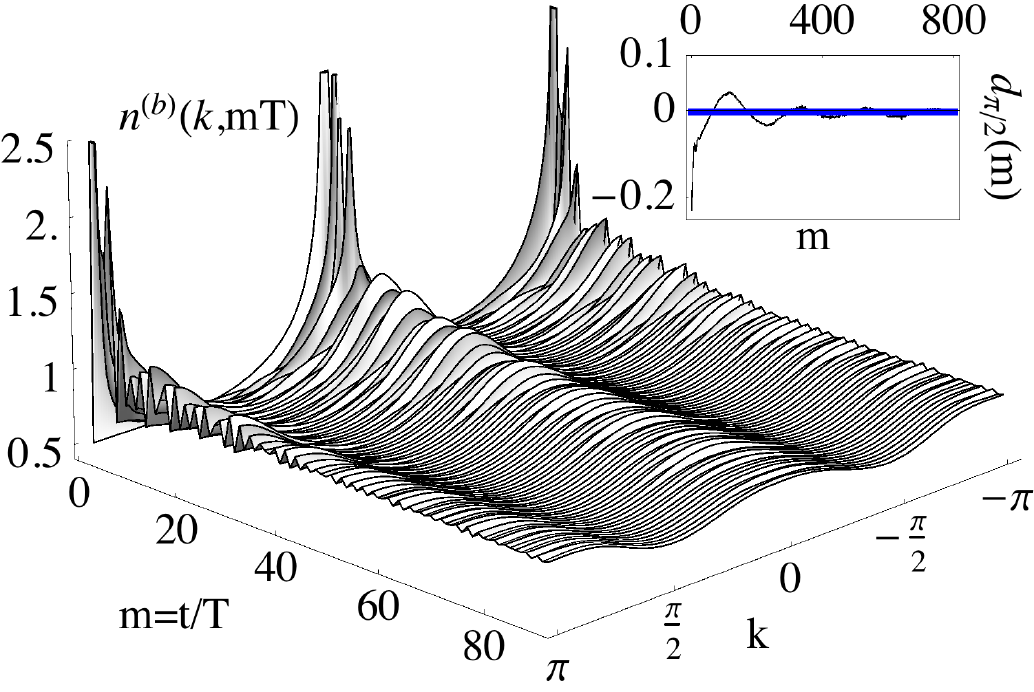}
  \end{center}
  \caption{
        {\bf Main plot:} Stroboscopic approach to 
		equilibrium with time for the full momentum
		distribution of the bosons, $\hat{n}^{(b)}$, corresponding to the heavy
		black line in Fig.~\ref{fig:final-Values-And-Period} and for
		a system size $L=200$ sites. Note the brief 
		initial transient period, followed by small oscillations around a well-defined
		limit.
		{\bf Inset:} Same as the main plot, but for
		a single component of the momentum distribution. 
		In this plot,
		$d_{\pi/2}(m)=\left(\hat{n}^{(b)}(k=\pi/2,mT)-\hat{n}_{PGE}^{(b)}(k=\pi/2)\right)/
			\hat{n}_{PGE}^{(b)}(k=\pi/2)$ 
		measures the deviation of the actual value from the prediction of the PGE. 
		The heavy blue lines show the average of the deviations 
		after discarding the first 50 periods, which approximates the long-time average. 
		These plots
		demonstrate that the expectation value of the operator 
		approaches, then oscillates about, a value that is very close (within a few percent) 
		to the PGE prediction. Both the deviation of the average from the PGE prediction
		and the relative magnitude of the fluctuations about the mean value 
		are shown to scale to zero with system size in Fig.~\ref{fig:approachtoeqlm}.
	  \label{fig:approach-equilibrium-full-nk}
}
\end{figure}

The main plot of Fig.~\ref{fig:approach-equilibrium-full-nk} 
shows the stroboscopic approach to the PGE state of
the full bosonic momentum distribution, $\hat{n}^{(b)}(k,mT)$, for the parameters corresponding
to the black line in Fig.~\ref{fig:final-Values-And-Period}. The entire momentum distribution
approaches, then oscillates around, a period-independent result. The inset focusses on the component $\hat{n}^{(b)}(k=\pi/2)$, showing the stroboscopic time
evolution of its difference from the value predicted by the PGE as a function of period, showing the oscillations about the equilibrium value shown by the heavy blue lines.

\begin{figure}[t]
  \begin{center}
    \includegraphics[width=8.5cm]{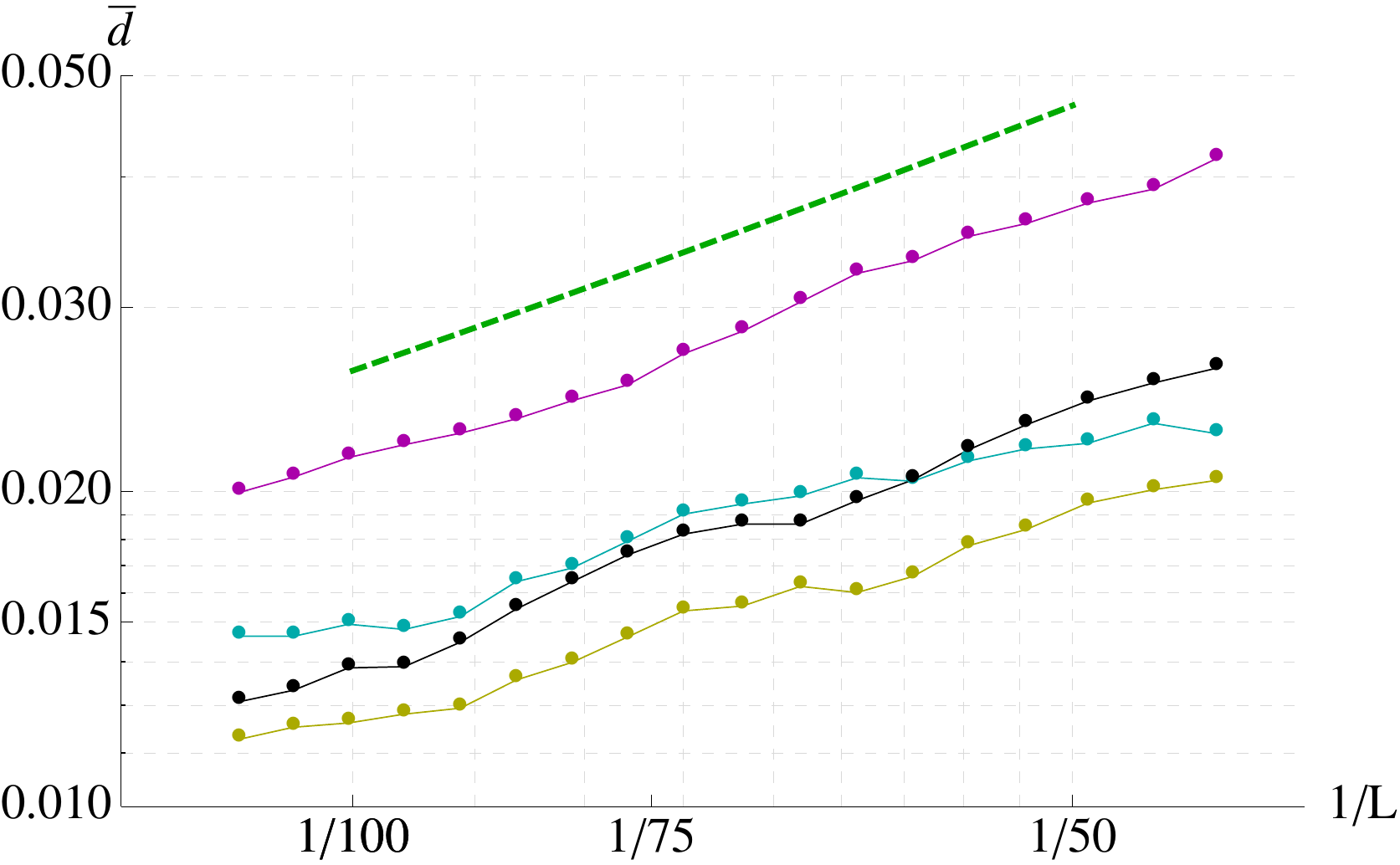}
  \end{center}
  \caption{(Color online) 
    Approach to equilibrium with system size. The Hamiltonian and colour coding
    is the same as in Fig.~\ref{fig:final-Values-And-Period}.
	Here, $\overline{d}$ measures distance from the PGE prediction,
    $\overline{d}=(LN)^{-1}\sum_{m=n}^{n+N}\sum_k \left|\hat{n}(k,mT)-\hat{n}_{PGE}(k)\right| $. 
	We take $n=40L$ and $N=20L$, large
    enough so that the results are insensitive to further increase. 
    The dashed green line is a plot of 
    $\overline{d}\propto L^{-1}$ to guide the eye. These results
    strongly suggest that the distance of the long-time behaviour of
    the system from our prediction at the thermodynamic limit falls
    off as a power law.}
  \label{fig:approachtoeqlm}
\end{figure}

We now quantitatively study the approach to the PGE
limit as system size is increased.
In Fig.~\ref{fig:approachtoeqlm} we plot
the average of the distance of the dynamical momentum distribution from its PGE value
over a number of periods, 
$\overline{d}=(LN)^{-1}\sum_{m=n}^{n+N}\sum_k
 \left|\hat{n}^{(b)}(k,mT)-\hat{n}_{PGE}(k)\right| $,
as a function of the inverse system size $1/L$.
These plots are for large $n=40L$ and $N=20L$ in order to
to allow plenty of time for equilibration. From 
Fig.~\ref{fig:approachtoeqlm}, we conclude
that the average of the momentum distribution
approaches the PGE result, while fluctuations away from it
average become smaller with increasing system size:
as $L\rightarrow\infty$, the momentum distribution rapidly
approaches the PGE periodic steady-state. 

In conclusion, we have shown that the real dynamics rapidly approaches the 
thermodynamic-limit and long-time results for relatively small systems and short times.

\emph{Experiments--}We now turn to the question of the experimental
implementation of the specific system we have studied. 
To realize our proposal, three ingredients
are required: A superlattice potential, periodic modulation and HCBs.

Experiments using a superlattice potential are already available~\cite{Atala2013}, while
periodic modulation of the lattice depth~\cite{StoferleModulation,Haller2010,StoferleModulation}
is a standard technique. In particular, periodically driving a superlattice potential is
described in Ref.~\cite{ChenModulation}. Finally, the HCB regime may be achieved via
confinement-induced resonance, which involves manipulating the radial harmonic potential
strength~\cite{Olshanii1998,haller2010confinement}. 

The example we have studied above is therefore accessible with current experimental techniques.

\emph{Conclusions and outlook--}For a large class of integrable periodically-driven 
systems, we have shown that a periodic steady-state is attained at long times. 
To describe this state, we have constructed a periodic version of the 
generalized Gibbs ensemble (GGE)~\cite{Rigol2007a}, commonly introduced in 
connection  with quenches in integrable models. We have provided an analytical demonstration 
that it exactly reproduces the periodic steady-state in the thermodynamic limit. 
We also provide numerical evidence of rapid convergence (i) to the 
thermodynamic-limit prediction with increasing system size and (ii) to the steady-state 
with time.

It would be natural to extend our results to a generic nonintegrable
situation. Our PGE is analogous to the GGE for non-driven systems~\cite{Rigol2007a};
the analogy would suggest that, for a closed, 
non-integrable, periodically-driven system, 
a subsystem for which the rest of the system plays the role of a bath might be described
by the periodic density matrix operator 
$\exp\left(\heff\left(\eps\right)=\hat U\mb\left(0,\eps\right)\heff \hat U\mb^\dagger\left(0,\eps\right)\right)$, 
analogous to the Gibbs ensemble for non-driven systems~\cite{Popescu2006}. 
Unfortunately, there are several 
issues with this; chief amongst them are that $\heff\left(\eps\right)$ is not a local 
operator in general and, more seriously, that $\heff\left(\eps\right)$ is not uniquely defined 
(its eigenvalues are only defined modulo $2\pi/T$--we do not use the eigenvalues
and therefore circumvent this problem in our work). We are currently investigating possible
resolutions of these conceptual issues.

Our work here should be compared to the usual situation for 
out-of-equilibrium systems, where each case has to be studied
individually using ad-hoc techniques tailored to the specific problem
at hand. In contrast, for this type of
periodically-driven systems the general framework of maximum entropy
statistical mechanics applies as-is.
It not only gives the correct ensemble but also
allows detailed computation of physical observables. We hope that this
work will motivate the search for further such ``thermodynamic''
principles governing driven systems in all generality.

\section{Acknowledgments} We acknowledge discussions with 
M.~Aidelsburger, M.~Atala and J.~T.~Barreiro. A.~L. thanks M. Kollar, O. Tieleman,
P. Ribeiro, A. Eckardt, T. Scheler, A. Sen, V. Bastidas, and M. Haque for discussions. AD acknowledges inspiring general discussions with E. Tosatti on non-equilibrium in the past.

%


\renewcommand{\bra}[1]{\left< #1 \right|}
\renewcommand{\ket}[1]{\left| #1 \right>}
\renewcommand{\heff}{\hat{H}_{\mathrm{eff}}}
\renewcommand{\tred}[1]{{\color{red} {#1}}}
\renewcommand{\tblue}[1]{{\color{blue} #1}}
\renewcommand{\eps}{\epsilon}
\renewcommand{\mI}[1]{\hat{\mathcal{I}}_{#1}}
\renewcommand{\U}[2]{\hat{U}(#1,#2)}
\renewcommand{\Ud}[2]{\hat{U}^\dagger(#1,#2)}
\renewcommand{\mA}{\hat{\mathcal{A}}}
\renewcommand{\tc}{\tilde{a}}
\renewcommand{\tb}{\tilde{b}}
\renewcommand{\hb}{\hat{b}}
\renewcommand{\hbd}{\hat{b}^\dagger}
\renewcommand{\hc}{\hat{c}}
\renewcommand{\ha}{\hat{a}}
\renewcommand{\hcd}{\hat{c}^\dagger}
\renewcommand{\hh}{\hat H}
\renewcommand{\mb}{}
\renewcommand{\cN}{\mathcal{N}}
\renewcommand{\cM}{\mathcal{M}}
\renewcommand{\hs}{\hat{\sigma}}

\title{Supplementary material for ``Periodic thermodynamics of isolated quantum systems''}
\author{Achilleas Lazarides$^{1}$, Arnab Das$^{1,2}$ and Roderich Moessner$^{1}$}
\affiliation{$^{1}$ Max-Planck-Institut f\"ur Physik komplexer Systeme, 01187 Dresden, Germany}
\affiliation{$^{2}$ Theoretical Physics Department, Indian Association for the Cultivation of Science, Kolkata 700032, India}
\date{\today}
\maketitle

\section{Synchronization} 
\label{sec:emergence_of_diagonal_ensemble} 
We first show how a periodically driven system synchronizes with the driving. This is 
analogous to the way a non-driven system equilibrates~\cite{Reimann2008a,Reimann2012a}, 
with most observables approaching a time-independent steady-state in which
contributions from off-diagonal (in the energy basis) matrix elements are negligible. 
This is usually called the ``diagonal ensemble'' (DE).

Consider the expectation value of the operator $\mathcal{\hat O}$ at time 
$t_n=\eps+nT$, 
$\mathcal{O}(t_n)=\mathrm{tr}\left(\hat{\rho(t_n)}\mathcal{\hat O}\right)$. Introducing the rotated operator defined in the main text,
\begin{equation*}
  \mathcal{O}(t_n)=
  \mathrm{tr}
  	\left(
    	\mathrm{e}^{-i\heff\mb nT}
		\hat{\rho}(0)
		\mathrm{e}^{i\heff\mb nT}
		\mathcal{\hat O^{(\epsilon)}}
	\right)
\end{equation*}
can be viewed as the expectation value of the rotated operator $\mathcal{\hat O^{(\epsilon)}}$ 
evolving under a time-independent Hamiltonian $\heff$ at time $nT$ starting
from the initial state $\hat{\rho}(0)$. 
For such a static quench, and under a set of general assumptions for the initial 
state~\cite{Reimann2008a,Reimann2012a,Cazalilla2012a}, 
one expects each series $\{{\mathcal O}(nT+\epsilon);\, n = 0,1,2\ldots\}$ to 
converge to a 
fixed value~\footnote{To be more precise, for a finite system, successive 
elements of $\{{\mathcal O}(nT+\epsilon; n = 0,1,2\ldots)\}$ 
will approach and then oscillate around a time-independent value.}.
Denote the eigenvalues and eigenstates of  
$\heff$ by  $\epsilon_{\alpha}$ and $\ket{\alpha}$, respectively,
with $\alpha=1\ldots D_{\mathcal{H}}$ and $D_{\mathcal{H}}$  
the dimension of the Hilbert space of the system.
Expanding $\hat{\rho}(0)=\sum_{\alpha,\beta} \rho_{\alpha,\beta}
			\ket{\alpha}\bra{\beta}$,
the limit of the long-time average over many periods is given by
\begin{equation}
  \label{eq:DE}
	  \lim_{N\rightarrow\infty}N^{-1}\sum_{n=1}^{N} \mathcal{O}(nT+\epsilon)
  =
  \sum_{\alpha=1}^{D_\mathcal{H}} \rho_{\alpha,\alpha}
  \left<\alpha\right|\hat{\mathcal{O}}^{(\eps)}\left|\alpha\right>,
\end{equation}
analogously to the DE result for a static system.

It has been shown by Reimann~\cite{Reimann2008a} that there are two necessary 
conditions for the equality~\eqref{eq:DE} to be accurate. Firstly, defining the inverse participation ratio in the eigenstate basis by 
\begin{equation}
	\phi_q=\sum_\alpha \left| c_\alpha \right|^{2q}
\end{equation}
it is necessary that $\phi_2\ll 1$, that is, a sufficiently large fraction 
of the eigenstates of $\heff$ 
must be occupied. Secondly, the range of possible eigenstate expectation values of the 
operator in question, $\mathcal{\hat{A}}=\hat{\mathcal{O}}^{(\eps)}$, must be finite, ie, 
$\Delta_{\mathcal{A}}=\max_\psi\bra{\psi}\mathcal{\hat{A}}\ket{\psi}
	-\min\bra{\psi}\mathcal{\hat{A}}\ket{\psi}$ must be finite. 
If these two conditions hold, then a modification of the arguments 
of Ref.~\cite{Reimann2008a,Reimann2012a} shows that the mean square deviation of the actual 
time evolution from the prediction of the diagonal ensemble,
\begin{equation}
  \label{eq:stdreimann}
	\sigma_\mathcal{A}^2=\overline{
		\left(
			\mathcal{O}( n T+\epsilon)-\mathrm{tr}
        \left(\hat{\mathcal{O}}\hat{\rho}_{PGE}(\eps)\right) 
		\right)^2
	}
\end{equation}
where $\overline{f(n)}=\lim_{N\rightarrow\infty}N^{-1}\sum_{n=1}^{N} f(n)$ is bounded by
\begin{equation}
  \label{eq:boundreimann}
	\sigma_\mathcal{A}^2\leq \Delta_{\mathcal{A}}\phi_2.
\end{equation}

All observables we consider (such as the single-particle momentum distribution) clearly 
have a finite $\Delta_\mathcal{A}$, so that we conclude that synchronization occurs for 
any initial state sufficiently nonlocal in the basis formed by the eigenstates of 
$\heff$. 

\section{Proof that the PGE captures the synchronized state}
\label{sec:correctness-pge}

We now turn to the special case of integrable systems that can be mapped
to the form of Eq. (3) of the main text (possibly via a Jordan-Wigner transformation). 
For simplicity, we also specialise to pure initial states, such that 
$\hat{\rho}(0)=\ket{\psi(0)}\bra{\psi(0)}$ and $\rho_{\alpha,\beta}=c_\alpha^*c_\beta$ 
with $c_\alpha^*=\left<\psi(0)\right.\left|\alpha\right>$.
Our goal is to show that the expectation value of any operator $\mathcal{\hat{O}}$ 
at any time $\eps$ in the long-time limit is equal 
to $\mathrm{tr}\left(\hat{\mathcal{O}}\hat{\rho}_{PGE}(\eps)\right)$.

In this section, the operators $\ha$ refer to the operators
diagonalising $\heff$ (they are defined in Eq. 4 of the main text), while the 
states denoted by Greek letters such as $\ket{\alpha}$ refer to the
\emph{many-body} eigenstates of $\heff$, as in the previous section.

Let us begin by considering bilinear operations $\hat{\mathcal{O}}$.
Writing $\mA=\hat{\mathcal{O}}^{(\eps)}$, defining the long-time limit 
(see Eq.~\eqref{eq:DE})
$\mathcal{A}_{L}=\sum_{\alpha=1}^{D_\mathcal{H}} \left|c_\alpha\right|^2 \left<\alpha\right|\mA\left|\alpha\right> $ 
and expanding $\mA$ in the $\tc_p$, 
$\mA=\sum_{p,q}\left(
    \mathcal{A}_{p,q}\tc^\dagger_p\tc_q+\mathcal{B}_{p,q}\tc_p\tc_q
    +\mathcal{C}_{p,q}\tc^\dagger_p\tc_q^\dagger
  \right)$, 
we have
\begin{equation}
  \mathcal{A}_{L} =\sum_{\alpha,p,q}\mathcal{A}_{p,q}\left|c_\alpha\right|^2\bra{\alpha}\tc^\dagger_p\tc_q\ket{\alpha}.
\end{equation}
We now use the identities
$\bra{\alpha}\tc^\dagger_p\tc_q\ket{\alpha}=
\bra{\alpha}\tc^\dagger_p\tc_q\ket{\alpha}\delta_{p,q}$ 
and
$\bra{\alpha}\tc^\dagger_p\tc_q\ket{\beta}\delta_{p,q}
	=\bra{\alpha}\tc^\dagger_p\tc_q\ket{\alpha}\delta_{p,q}\delta_{\alpha,\beta}$, 
	the second of which follows from 
	$\left[\heff,\hat{N}\right]=0$ with $\hat{N}$ the 
  particle number operator, finally arriving at
\begin{equation}
  \mathcal{A}_{L}=\sum_{p}\mathcal{A}_{p,p}\mathcal{I}{p}(0)
\end{equation}
with $\mathcal{I}{p}(0)=\bra{\psi(0)}\mI{p}(0)\ket{\psi(0)}
	=\bra{\psi(0)}\tc^\dagger_p\tc_p\ket{\psi(0)}$. Therefore,
\begin{equation}   
  \mathcal{A}_{L}=
    \sum_{p=1}^L\mathcal{A}_{pp}
      \bra{\psi(0)}\hat{\mathcal{I}}_p(0)\ket{\psi(0)}.   
\label{eq:de-eq-pge} 
\end{equation} 
Explicit calculation then shows that 
$\mathrm{tr}\left(\hat{\mathcal{O}}\hat{\rho}_{PGE}(\eps)\right)=
\mathrm{tr}\left(\hat{\mathcal{A}}\hat{\rho}_{PGE}(0)\right)=\mathcal{A}_{L}$. 
Thus, the long-time limit for the expectation of any bilinear
operator $\mA$ is exactly reproduced by the PGE density matrix
for any time $\eps\in[0,T)$.

Using Wick's theorem we can extend this result to 
higher-order (than bilinear) operators, at least in the absence of accidental 
symmetries~\cite{Fagotti2013,Dora2012a} and in the thermodynamic limit (both conditions
are necessary in order for fluctuations to vanish and Wick's theorem to be applicable). 

Consider an arbitrary higher-order operator 
$\hat{\mathcal{O}}$, involving terms 
with more than two fermionic operators (it is always possible to express any
$\hat{\mathcal{O}}$ as a sum of products of fermionic operators). 

We now take an initial state that is an eigenstate of some bilinear Hamiltonian without 
translational invariance~\footnote{Translational invariance introduces ``accidental'' 
correlations in the initial state~\cite{Fagotti2013,Dora2012a}}. As the time 
evolution occurs under another quadratic Hamiltonian, 
the expectation value of each product of fermionic operators factorises at each instant in 
time according to Wick's theorem~\cite[Sec. 14]{StatPhys2}. For example, 
for an $M$-particle operator, 
$\hat{\mathcal{A}}=\tc^\dagger_{p_1}\cdots\tc^\dagger_{p_M}\tc_{p_1}\cdots\
\tc_{p_M}$, defining $\left<\cdots\right>(t)=\mathrm{tr}\left(\cdots \hat{\rho}(t)\right)$ 
we have 
$\left<\hat{\mathcal{O}}\right>(t)=
  \sum_{\mathcal{P}}(-1)^\mathcal{P}
    \left<\tc^\dagger_{p_{\mathcal{P}_1}}\tc_{p_{\mathcal{P}_1}}\right>(t)\cdots
    \left<\tc^\dagger_{p_{\mathcal{P}_M}}\tc_{p_{\mathcal{P}_M}}\right>(t)$ 
    where $\mathcal{P}$ denotes a permutation. 
\emph{If} the long-time limit
$\lim_{t\rightarrow\infty}\left<\tc^\dagger_p \tc_q\right>(t)$ 
exists, then, by the earlier argument for bilinear operators, it is given 
by the PGE result and therefore 
$\lim_{N\rightarrow\infty}N^{-1}\sum_{n=0}^N\left<\mathcal{O}\right>(nT+\eps)=\mathrm{tr}\left<\mathcal{O}\hat{\rho}_{PGE}(\eps)\right>$ even for the higher-order correlators. 
If, on the other hand, the limit does not exist then the proof fails and higher-order operators are not guaranteed to be reproduced by the PGE. 
For infinite systems, $\left<\tc^\dagger_p \tc_q\right>(t)$ generally approaches a limit for $t\rightarrow\infty$

In conclusion, the main assumption necessary for this result is that the expectation values
of bilinear operators tend to a well-defined limit at long times; this is generally true for
systems in the thermodynamic limit~\cite{Reimann2008a,Reimann2012a}. However, even in the 
thermodynamic limit it is known to fail for disordered systems~\cite{Ziraldo2012a,Ziraldo2013a}.

\section{Example Hamiltonians} 
\label{sec:example_hamiltonians}

In the main text we concentrate on a system of hard-core bosons.
Here, we explicitly list a number of other important
physical Hamiltonians that may be mapped to the form
\begin{equation}
  \hat{H}(t) = \sum_i\left[
      \ha_i^\dagger \cM_{i,j}(t)\ha_j + 
      \ha_i^\dagger \cN_{i,j}(t) \ha_j^\dagger + \mathrm{h.c.}
    \right],
  \label{eq:quadratic}
\end{equation}
with $\ha$ either bosonic or fermionic.

{\bf Luttinger Liquids:}
Another important class of Hamiltonians with 
broad applications is given by Eq.~\eqref{eq:quadratic} with 
the $\ha$ satisfying bosonic commutation relations. 
In particular, Luttinger liquids 
(LLs)~\cite{Haldane1981a,GiamarchiBook} are in the class of one-dimensional systems 
described by such a Hamiltonian.
As a concrete example, the Hamiltonian for a spatially homogeneous 
time-dependent LL may be written in the 
  form~\cite{Cazalilla2004a,Dora2011a,Dora2012a} 
  \begin{equation}
    \hat{H}=\sum_{q\neq 0}\left( \omega(q,t) \hb^\dagger_q \hb_q
      +\frac{1}{2}g(q,t)\left[ \hb_q \hb_{-q} + \hb_q^\dagger \hb_{-q}^\dagger \right] \right)
      \label{eq:hLL-homogeneous}
  \end{equation}
with $\hb_q$ bosonic operators and $\omega(q,t)$ and $g(q,t)$ periodically 
time-dependent coefficients~\footnote{
  Physically, a LL is usually obtained as an approximate hydrodynamic description of, 
  for example, a strongly-interacting bosonic system in one
  dimension~\cite{Haldane1981a,GiamarchiBook,Cazalilla2004a,Gangardt2003,Lazarides2011a}, 
  and is not expected to correctly describe highly excited states of the physical system. 
  The question of whether the dynamics of a given periodically-driven system 
  is correctly described by a driven LL therefore requires a detailed 
  case-by-case analysis.
  Here, we consider a periodically-driven LL as a model system
  without discussing its applicability to specific experimental
  situations.
}.

{\bf XY Chain:}
Another very well-studied model Hamiltonian is the spin-1/2 quantum $XY$ chain, for which 
$\hat{H}_{XY}(t)=\sum_i\left[ J_x (t) \hs^x_i \hs^x_{i+1}
  + J_y (t) \hs^y_i \hs^y_{i+1} 
  +B(t)\hs^z_i\right]$ with the $\hs$ spin-1/2 operators. Using again a standard 
  Jordan-Wigner transformation~\cite{WenBook}, this is mapped to 
  Eq.~\eqref{eq:quadratic} with 
  $\cM_{i,j}(t)=\left(J_x(t)+J_y(t)\right)\left(\delta_{i,i+1}+\delta_{i,i-1}\right)+
   B(t)\delta_{i,j}$
  and
$\cN_{i,j}(t)=\left(J_x(t)-J_y(t)\right)\left(\delta_{i,i+1}+\delta_{i,i-1}\right)$.

\end{document}